\documentclass[aps,prl,twocolumn,superscriptaddress,floatfix,amsmath]{revtex4}
\usepackage{graphicx}
\begin{document}

\today
\pacs{}
\title{From particles to spins: 
Eulerian formulation of supercooled liquids and glasses}
\author{Claudio Chamon}
\address{Physics Department, Boston University, Boston, MA 02215, USA}
\author{Leticia F. Cugliandolo}
\address{Universit\'e Pierre et Marie 
Curie -- Paris VI, LPTHE UMR 7589, 4 Place Jussieu,  75252 Paris Cedex 05, 
France}
\author{Gabriel Fabricius}
\address{Instituto de Investigaciones Fisicoqu\'imicas 
Te\'oricas y Aplicadas (INIFTA), UNLP, CONICET. Casilla de Correo 16, 
Sucursal 4, (1900) La Plata, Argentina},
\author{Jos\'e Luis Iguain}
\address{Departamento de F\'isica, FCEyN, 
Universidad Nacional de Mar del Plata, De\'an Funes 3350, 7600 
Mar del Plata, Argentina},
\author{Eric R. Weeks}
\address{Physics Department, Emory University, Atlanta, GA 30322, USA}

\begin{abstract} 
The dynamics of supercooled liquid and glassy systems are usually
studied within the Lagrangian representation, in which the positions
and velocities of distinguishable interacting particles are
followed. Within this representation, however, it is difficult to
define measures of spatial heterogeneities in the dynamics, as
particles move in and out of any one given region within long enough
times. It is also non-transparent how to make connections between the
structural glass and the spin glass problems within the Lagrangian
formulation. We propose an Eulerian formulation of supercooled liquids
and glasses that allows for a simple connection between particle and
spin systems, and that permits the study of dynamical heterogeneities
within a fixed frame of reference similar to the one used for spin
glasses. We apply this framework to the study of the dynamics of
colloidal particle suspensions for packing fractions corresponding to
the supercooled and glassy regimes, which are probed via confocal
microscopy.
\end{abstract}

\maketitle

\section{Introduction}

The phenomenology of structural and spin glasses
has much in common: no static long-range order, aging relaxation,
heterogeneous dynamics, etc~\cite{Cu}. While a precise and unambiguous
connection between these two problems still lacks, the possibility
that such relation exists dates back to the work by Kirkpatrick,
Thirumalai and Wolynes~\cite{KTW1,KTW2,KTW3}, who proposed a
connection between structural glasses and the p-spin disordered
model. More recently, Tarzia and Moore~\cite{Tarzia-Moore} have
paralleled the phenomenology of structural glasses to that of an
Edwards-Anderson model in a uniform magnetic field. One of the main
hurdles in making a direct real space connection between these two
problems is that spin glass models are defined on a lattice, while the
particles comprising structural glasses are itinerant.

Supercooled liquids and glasses are usually described within the
Lagrangian formulation, where one tracks the position of individual
particles as a function of time. Natural quantities computed within
this frame of reference are mean-square displacement and
self-diffusion of these particles. Heterogeneous dynamics can be
probed, for example, by studying quantities such as mobility within
prescribed boxes; however, such fixed regions serve this purpose just for 
a certain time, as particles move in and out of these boxes if one
waits for long enough.  In contrast, studying local dynamics in a spin
glass presents no such complication, as spins remain fixed to their
sites at all times, and all that changes is the spin orientation as
function of time. Therefore, if one is to construct a simple
description of particle systems that could actually be used in
analyzing real experimental data from the point of view of a spin
glass, one must abandon the Lagrangian formulation.

We propose an Eulerian analysis of the dynamics of interacting
particle systems. The proposal consists in transforming the data of
numerical simulations or confocal microscopy experiments, usually
presented in the Lagrangian representation as time-dependent positions
and velocities of distinguishable particles~\cite{Weeks,Cicowe,Wecrwe,Makse}, 
into time-dependent
occupation numbers of finite volume pixels at fixed positions in
space. In this way, we divide the simulation or experimental fixed
volume box containing the particles into pixels located at sites
labeled by $i=1,\dots,N$, and assign a `spin', $s_i=\pm 1$, depending
on whether a `piece' of the particle falls within the pixel or
not. The data treatment here proposed allows one to make close contact
with spin (disordered or constrained) models.

\section{Method}

We divide the experimental box in cubic pixels of linear size $a=R/q$
where $R$ is the radius of the particles and $q$ is a parameter,
typically with $q>1$. The number of pixels is then $N=V/a^d$ with $V$
the total volume of the experimental box (we focus throughout the
paper on the $d=3$ case relevant to the experiments analyzed
below). The simplest definition of the spin variable is such that
$s_i=1$ whenever a particle (independently of which one it is)
overlaps the $i$-th pixel, and $s_i=-1$ otherwise. With such a
definition, though, the magnetization density is non-zero, $m=N^{-1}
\sum_{i=1}^N s_i \neq 0$, at a generic volume fraction $\phi$. To work
at zero magnetization density and make closer contact with usual spin
(glass) problems, we shrink the particle size to an effective radius
$R_{\rm eff}$ such that the covered volume is $50\%$~\cite{foot}.

An efficient algorithm that maps particle positions into spin
variables works as follows. First, one constructs the grid of pixels
and sets all spins to $s_i=-1$ for all $i$. Next, one reads the
particle centers from the data file and sets to $+1$ the spin variables
of the pixels around the center of each particle. One thus avoids
having to go over all sites in the lattice and to compute distances
between particle centers. This procedure is repeated at each time
step.

The data set has now been transformed into spin values and all
interesting correlation functions in the spin realization inform us
about the dynamics of the particle system. The spin variable is
naturally related to an occupation number, by $n_i\equiv
(s_i+1)/2=0,1$, and then to a density. We stress here that, in this
construction, these densities are not coarse grained quantities built
by looking at distances larger than the particle size, but instead the
other way around, by looking at distances of the order and below the
particle size. Within this construction, the parallel with the
spin-glass problem is also clear: a short-ranged equal-time spin-spin
correlation function corresponds to a short-ranged particle density
order, etc.

\begin{figure}
\includegraphics[width=8cm]{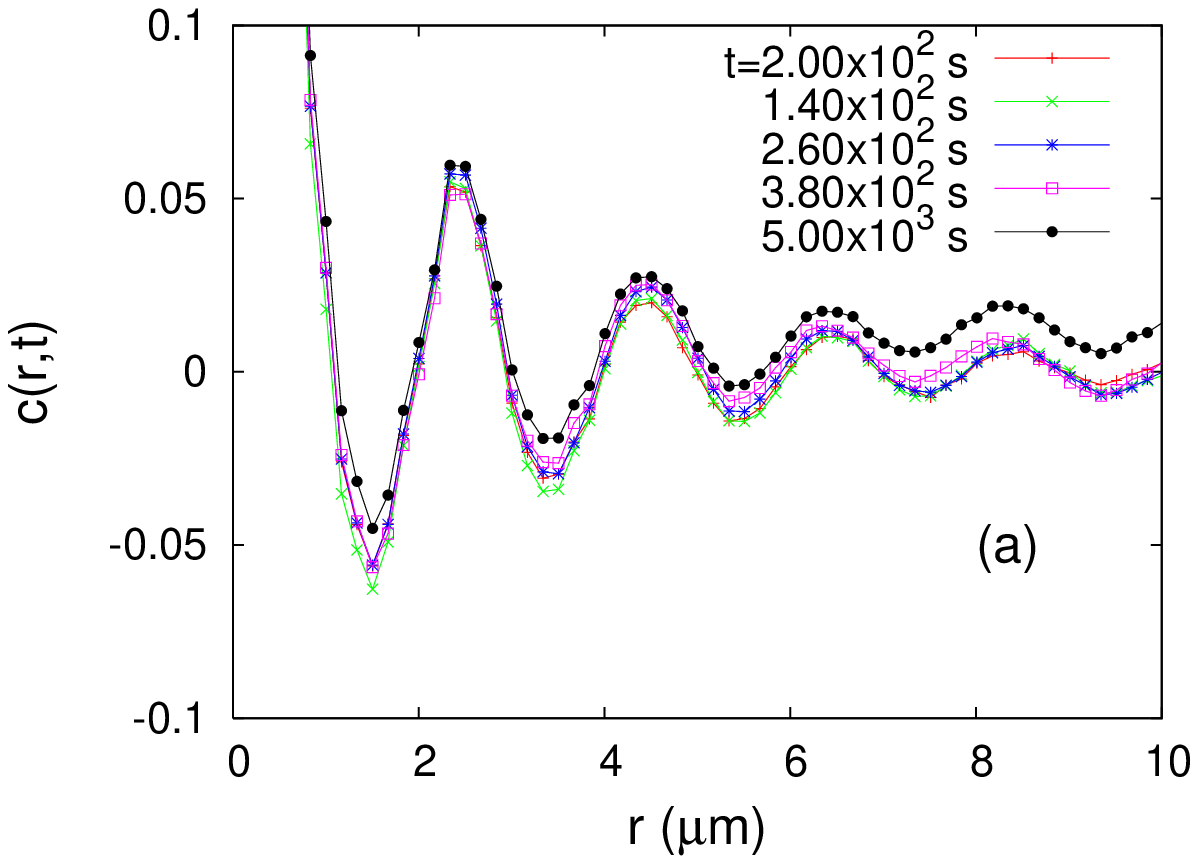}
\includegraphics[width=8cm]{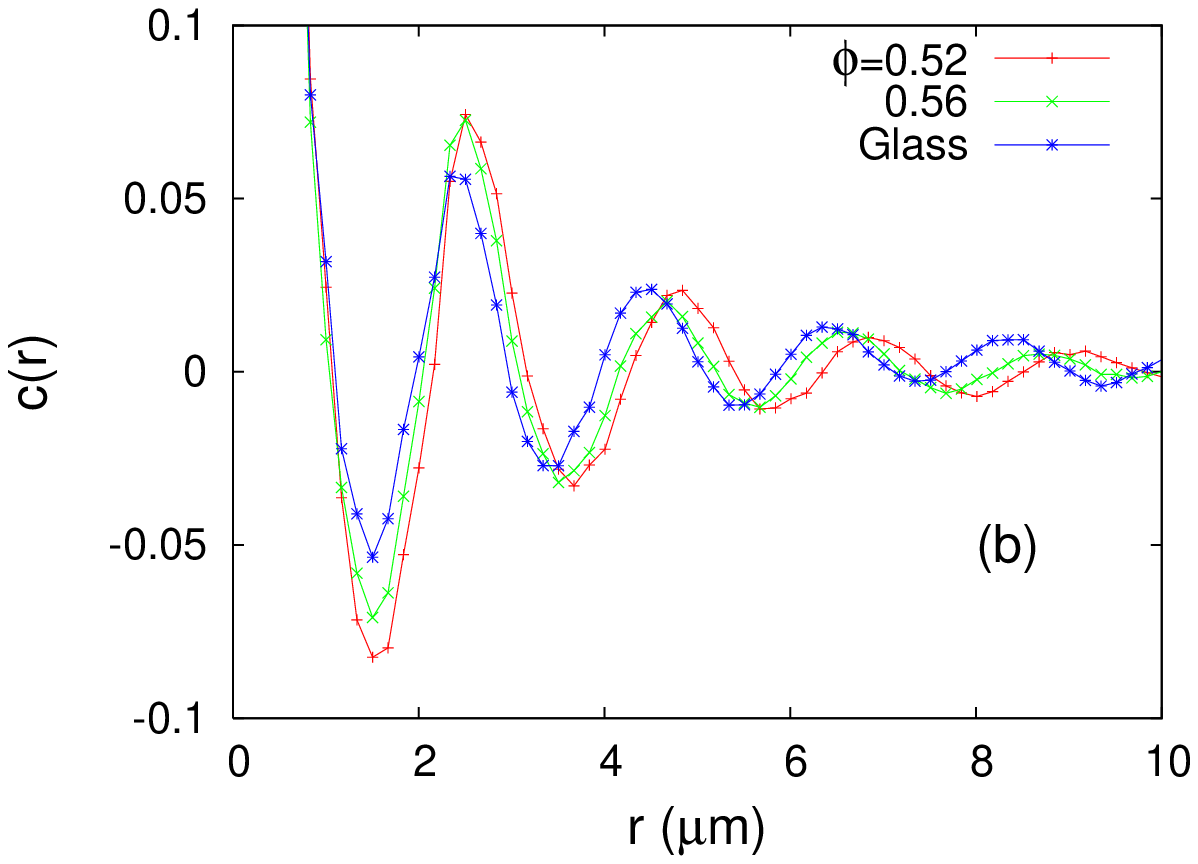}
\caption{\label{fig:1} (Color online) One-time two-spin correlation
  $c(r,t)$ as a function of distance $r$.  Panel (a): evolution with
  time of the glass ($\phi\simeq 0.62$) pair-correlation function. All
  curves are very similar, apart from the last one that does not decay
  at long distances. See the text for a discussion on this fact.
  Panel (b): the time-averaged quantity $c(r)$ for the three packing
  fractions, $\phi\simeq 0.52,\;0.56$, and $0.62$.}
\end{figure}

\section{Analysis}

We apply this framework to experimental data from colloidal
suspensions, both in the supercooled liquid regime~\cite{Weeks,Wecrwe}
and in the dense glassy phase~\cite{Cicowe}.  The suspensions are of
colloidal poly-methylmethacrylate (PMMA) with radius $R= 1.18$~$\mu$m
(and a polydispersity of $\sim 5\%$),
suspended in a mixture of decalin and either cycloheptylbromide (for
the samples with $\phi<\phi_g \approx 0.58$) or cyclohexylbromide (for
the sample with $\phi > \phi_g$). These solvent mixtures match the
index of refraction of the particles to aid in visualization.
Furthermore, the solvent mixtures also match the particle density, so
that sedimentation does not occur during the experiments.  In these
solvents, the particles are slightly charged, modifying their pair
correlation function somewhat from that of hard spheres, although they
still undergo the glass transition at $\phi_g \approx 0.58$.  The
particles in dilute samples diffuse their own diameter in $11\ s$,
although in these concentrated samples their motion is much slower
\cite{Weeks}.  All samples are stirred prior to data acquisition.  The
two samples with $\phi < \phi_g$ are stirred to break up any crystals,
and data acquisition is started after transient flows have diminished
($\sim 30$ min).  For the sample with $\phi > \phi_g$, no crystals are
present prior to stirring; instead, stirring initiates aging in the
sample, and data acquisition begins immediately after the stirring is
ended, setting the initial time $t_w=0$~\cite{Cicowe}.

\begin{figure}
\begin{center}
\includegraphics[width=8cm]{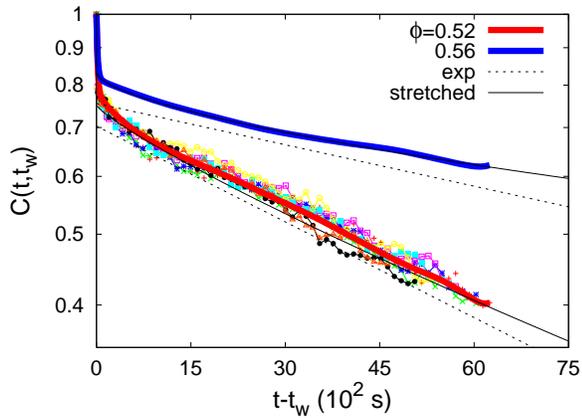}
\caption{
\label{fig:globalC} 
(Color online) Two-time two-spin local correlation $C(t,t_w)$ as a
function of time-delay $t-t_w$ in the supercooled liquids with
$\phi \simeq 0.52, \ 0.56$. The decay at several waiting-times ($t_w=180 \ s,
\ 306 \ s, \ 414 \ s, \ 558 \ s, \ 756 \ s, \ 1008 \ s, \ 1350 \ s, \
1800 \ s$) is shown with data points connected with thin lines; the
average of these sets is shown with thick lines, along with
exponential and stretched exponential fits to the averaged data.}
\end{center}
\end{figure}

Data is obtained via confocal microscopy \cite{Prasad}, which is used
to rapidly obtain a three-dimensional image of dimensions
approximately $60 \times 60 \times 12$~$\mu$m$^3$.  Within each image,
particle positions are obtained with an accuracy of 30 nm in $x$ and
$y$, and $50$ nm in $z$ (along the optical axis of the microscope).
For other experimental details, see Refs.~\cite{Weeks,Cicowe}.

\begin{figure}
\begin{center}
\includegraphics[width=8cm]{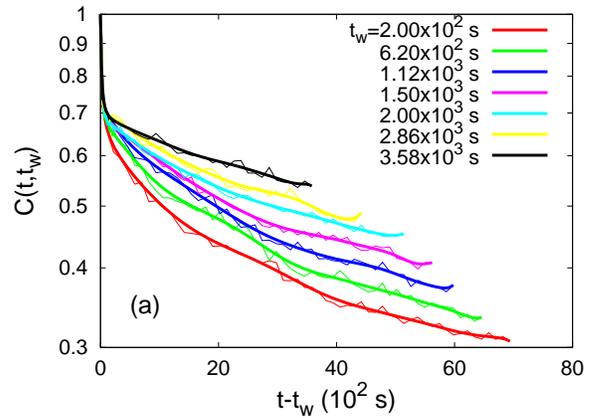}
\end{center}
\begin{center}
\includegraphics[width=8cm]{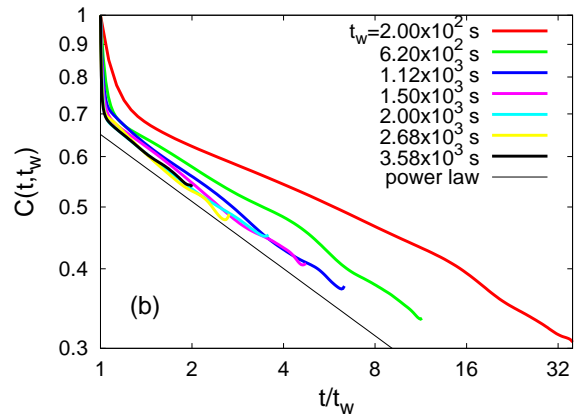}
\caption{
\label{fig:aging-decay} 
(Color online) Two-time two-spin local correlation $C(t,t_w)$ in the
glass. Panel (a): data for the decay at several waiting-times
($t_w=180\ s, \ 306\ s, \ 414\ s, \ 558\ s, \ 756\ s, \ 1008\ s, \
1350\ s, \ 1800\ s$) are shown with thin lines, plotted as function of
the time delay $t-t_w$; the smoothed decay is highlighted with thick
lines. Panel (b): scaled data using the simple aging form
$C(t,t_w)\sim f(t/t_w)$; the solid (black) line is the power law
$f(x)\sim x^{-0.35}$.}
\end{center}
\end{figure}

We choose $q=5$ and thus $a=R/q \simeq 1.18 \ \mu$m$/5\simeq 
0.236 \ \mu$m. The $3d$ positions of the particles were recorded every $18\, s$
for the supercooled data sets at $\phi\simeq 0.52$ and $\phi
\simeq 0.56$, and $20 \, s$ for the glassy data set at $\phi\simeq 0.62$.
For the effective radii we find: $R_{\rm eff}=1.17 \ \mu$m at $\phi\simeq 0.52$, 
$R_{\rm eff}=1.11 \ \mu$m at $\phi\simeq 0.56$
and $R_{\rm eff}=1.10 \ \mu$m at $\phi\simeq 0.62$.

We now show how to characterize the dynamics of the colloidal system
of particles using solely the mapped spin variables. We start by
defining two-spin correlations
\begin{equation}
C_2(r;t,t_w)=\frac{1}{N} \sum_{i,j;|\vec r_i-\vec r_j|=r}
s_i(t) s_j(t_w)
\;,
\end{equation}
which can be used to determine both equal-time spatial correlations
and same-site two-time correlations.

\begin{figure}
\begin{center}
\includegraphics[width=8cm]{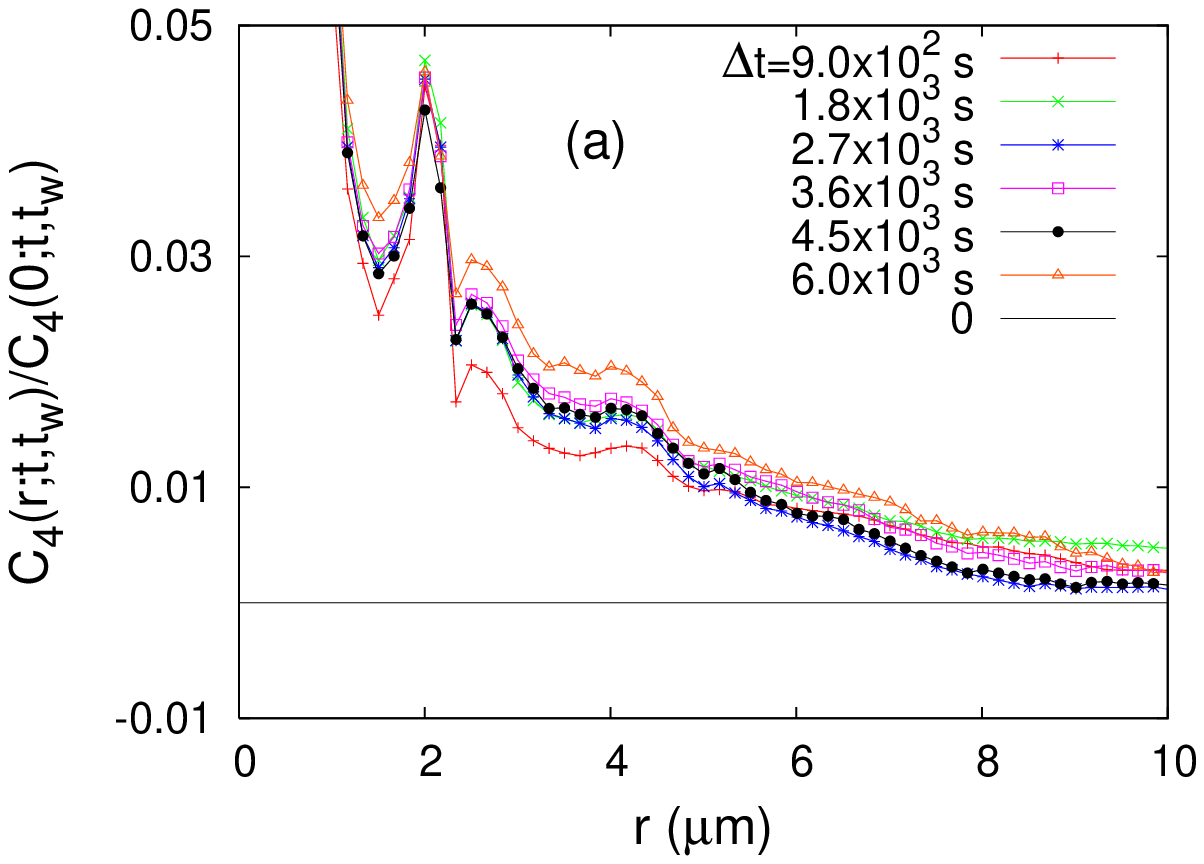}
\end{center}
\begin{center}
\includegraphics[width=8cm]{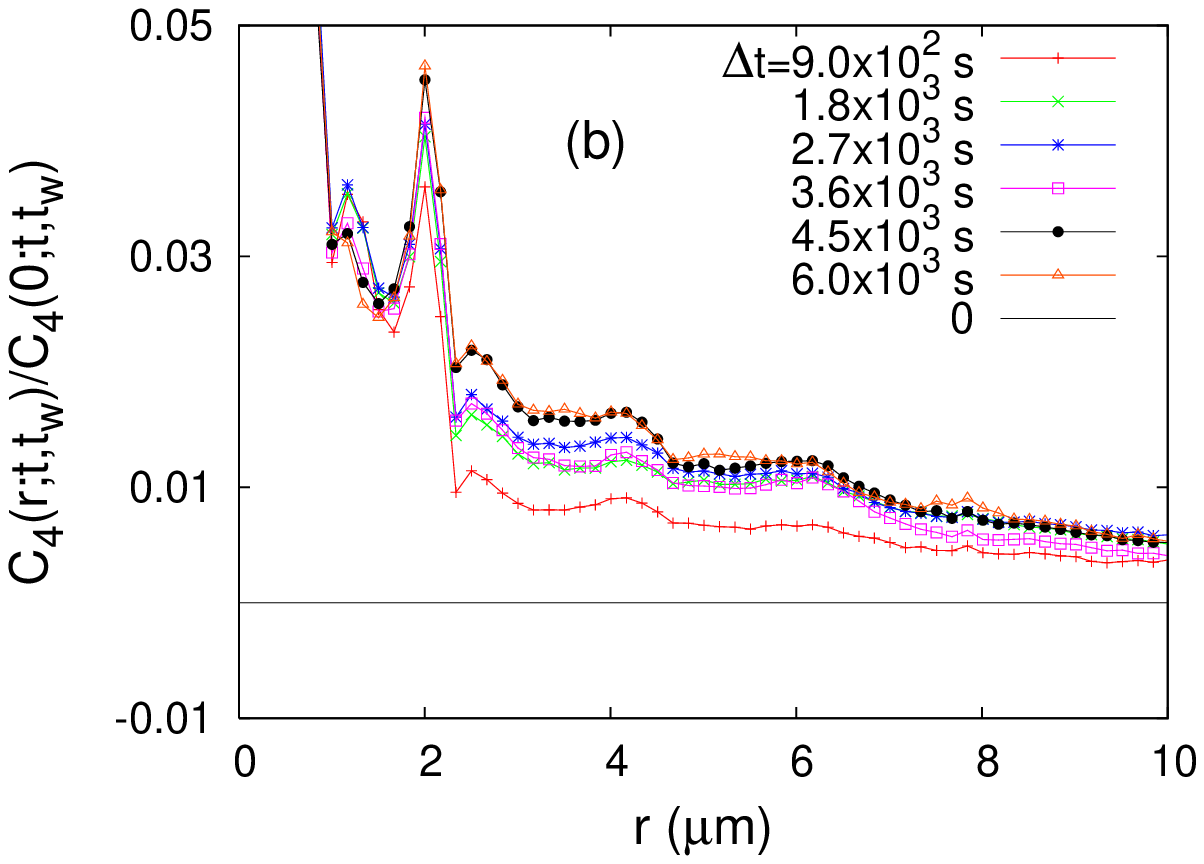}
\caption{\label{fig:C4-Deltat} (Colour online) The four point
correlation $C_4$ in the supercooled liquid for several time delays
given in the key.  Panel (a) $\phi\simeq 0.52$ and panel (b) $\phi\simeq 0.56$.}
\end{center}
\end{figure}

\begin{figure}
\begin{center}
\includegraphics[width=8cm]{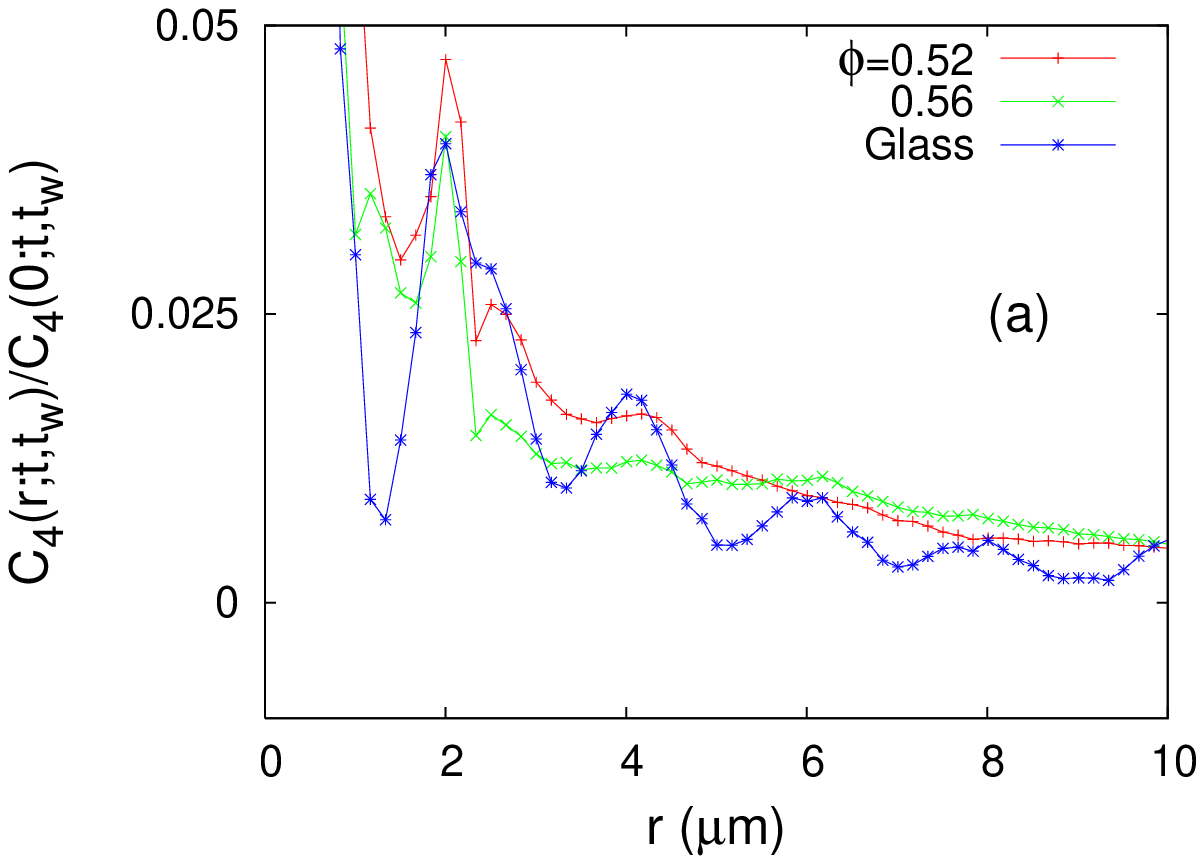}
\end{center}
\begin{center}
\includegraphics[width=225pt]{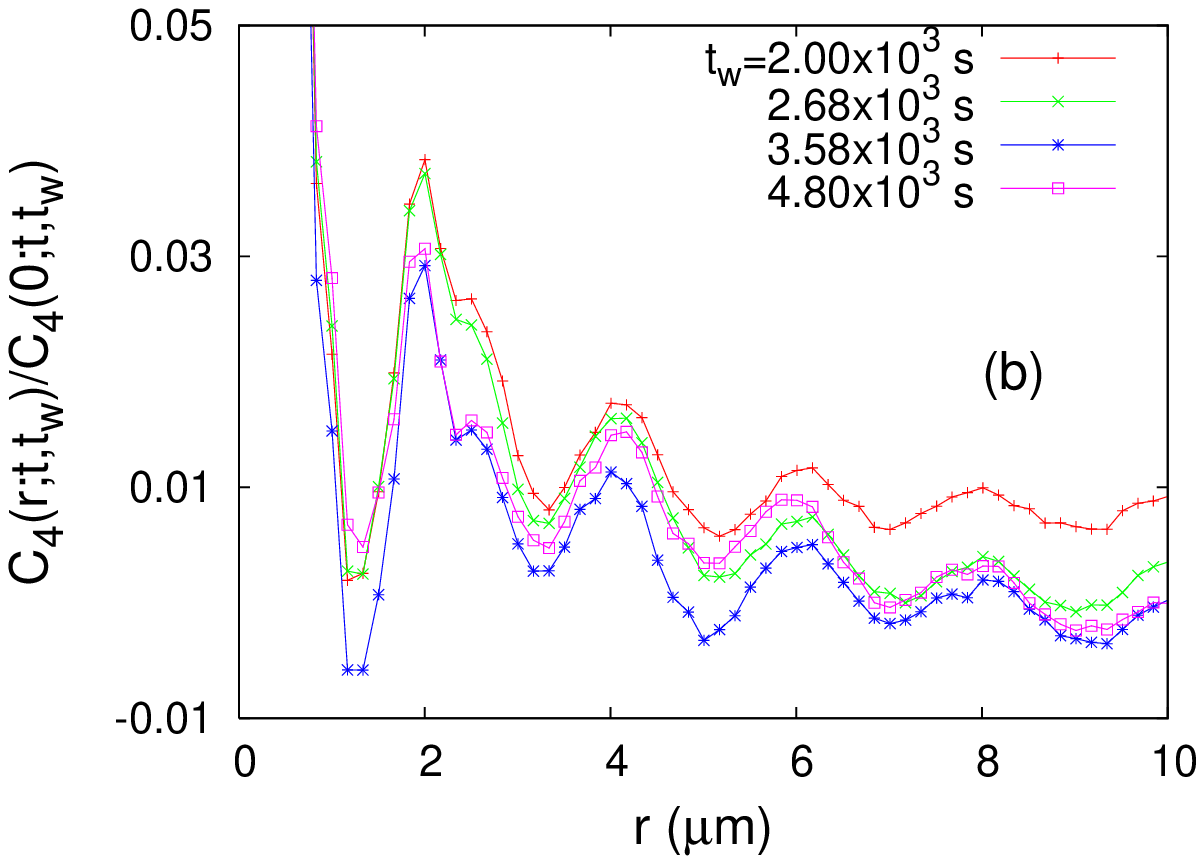}
\caption{\label{fig:C4} (Colour online) The four point correlation
$C_4$ in the supercooled liquid and the glass.  Panel (a): comparison
between three packing fractions, $\phi\simeq 0.52, \ 0.56$ at $t-t_w=1800\ s$
and the glass $\phi\simeq 0.62$ at $t-t_w=2000\ s$ and $t_w=2000\ s$. Panel (b):
$C_4$ for the glass at several waiting-times and $t-t_w=1200\ s$ fixed.}
\end{center}
\end{figure}

In Fig.~\ref{fig:1} we present the equal-time correlation function
between two spins at a distance $r$, $c(r,t)\equiv C_2(r;t,t)$, which
is analogous to the pair correlation function convolved with a square
hat function of width $R_{\rm eff}$.  Due to the finite size of the sample
one expects time-dependent fluctuations.  In the supercooled liquid
regime these are present but no systematic trend is visible (not
shown).  The time-dependence in the glass is shown in panel (a) where
the pair correlation function as a function of $r$, computed at
equally spaced times, is displayed.  The curves show no systematic
time-dependence until $t\sim 4000 \ s$. A clear departure is seen at
later times when the pair correlation no longer decays to zero. The
saturation at long-distances is not related to crystallization since
Cianci {\it et al} found no increase in crystalline order 
as the sample aged~\cite{Cicowe}.  We do not know the exact
reason for this saturation. In what follows we just analyze glass data
for times that are shorter than $t\sim 4000 \ s$.

The time averages of the equal-time correlation function, $c(r) \equiv
k_m^{-1} \sum_{k=1}^{k_m} c(r,t_k)$, are shown in panel~(b) of
Fig.~\ref{fig:1} for the three packing fractions.  These are
calculated using $k_m=10$ times equally spaced over an interval of
approximately $6300 \ s$ in the supercooled liquid and 8 times before
$4000s$ in the glass.  Notice that the peaks move slightly to lower
values of $r$ for increasing values of $\phi$, but there in no
qualitative difference in this one-time quantity for these three
packing ratios. These curves are essentially the same as the ones
shown in \cite{Wecrwe} for the supercooled liquid and in \cite{Cicowe}
for the glass, computed using directly the particle positions.

We now turn to two-time quantities, starting from the global
equal-space two-spin correlation $C(t,t_w)\equiv C_2(r=0;t,t_w)$. In
Fig.~\ref{fig:globalC} we present its decay, as a function of $t-t_w$
for the supercooled liquid regime.  The group of curves that fall
below are for $\phi\simeq 0.52$.  The curves drawn with thin lines represent
data for several waiting-times and, within the numerical error, they
have the same decay, proving that the dynamics are stationary. The
thick (red) line is the average over all waiting-times. The thick
(blue) curve lying above is the averaged data for $\phi\simeq 0.56$. The
spreading for different waiting-times (not shown) is similar to the
one for $\phi\simeq 0.52$.  In both cases $C$ decays from $1$ to $0.8$ in
less than $18\ s$ (the minimum time step for which data is recorded),
due to Brownian motion of the particles within their cages
\cite{Wecrwe}.  The dotted black lines are exponential fits, $f(x)=a
e^{-x/b}$, to the decay for $t-t_w>90\ s$ that have been translated to
make the curve visible. The solid black line is a fit of the data for
$\phi\simeq 0.56$ to a stretched exponential, $g(x)=c e^{-(x/\epsilon)^d}$,
fit to the averaged curves. In Table~\ref{tab:table2} we give the
values of the fitting parameters for both densities although in
Fig.~\ref{fig:globalC} we show only the stretched exponential for the
higher packing fraction.

\begin{table}
\caption{\label{tab:table2} 
Fitting parameters for global two-time correlation
decay in the supercooled liquid. The errors (not quoted)
are, at most, $3\%$.}
\begin{tabular}{ccccccc}
\hline\hline
$\phi$ & $a$ &$b\ (s)$&$c$& $d$ &$\epsilon\ (s)$\\
\hline
$0.52$ &
$0.75$ & $9950$ 
& $0.75$ & $0.85$ & $10650$ 
\\
$0.56$ & 
$0.80$ & $23000$ 
& $0.85$  & $0.65$ & $43400$\\
\hline 
\end{tabular}
\end{table}

Figure~\ref{fig:aging-decay} shows the two-time correlation function
in the aging regime. Panel (a) displays the relaxation after several
waiting-times. As in the supercooled liquid regime, the correlation
decays from $1$ to $0.7$ rapidly and then further decays to zero in a
much slower manner. In a double logarithmic scale the separation
between the stationary ($C\stackrel{>}{\sim} 0.7$) and aging ($C
\stackrel{<}{\sim}0.7$) regimes is seen as a plateau at the
Edwards-Anderson value $q_{eq} \sim 0.7$.  Panel (b) demonstrates that
the aging data can be satisfactorily scaled using the `simple' aging
form $C(t,t_w) \sim f(t/t_w)$ with $f(x)\sim x^{-0.35}$ for
waiting-times that are longer than $t_w\sim 1200\ s$.  However, the
range of variation of both axis is smaller than a decade and it is
hard to give a concrete conclusion on `simple' aging in this sense.
Still, it is interesting to note that this behavior is remarkably similar
to the one found with Monte Carlo simulations of the $3d$ Edwards-Anderson 
(EA) spin-glass~\cite{3dEA}.

A two-time dependent correlation length~\cite{Cachcuigke,Chcu} can be
extracted from the spatial decay of a two space points and two-times
correlation function:
\begin{eqnarray}
S_4(r; t, t_w) &=& 
\frac{1}{N} 
\sum_{i,j;|\vec r_i-\vec r_j|=r} 
s_i(t) s_i(t_w) s_j(t) s_j(t_w)
\; , 
\
\label{eq:S4}
\end{eqnarray}
or a variation in which we extract the square of the two-time local
correlation $C(t,t_w)$ that is the expected large distance limit of
eq.~(\ref{eq:S4}):
\begin{eqnarray}
C_4(r; t, t_w) &\equiv& S_4(r; t, t_w) - [C(t, t_w)]^2 
\; . 
\end{eqnarray}
These definitions are simple extension of the ones used in the analysis of the 
stationary supercooled liquid~\cite{Four-point,Four-point-MCT}.


Figure~\ref{fig:C4-Deltat} displays the four-point correlation
function for several time-delays in the supercooled liquid phase.  In
Fig.~\ref{fig:C4} we show the space-dependence of the four-point
correlation function $C_4$ for a fixed time-delay, $t-t_w=1800\ s$ for
the supercooled liquid and $t-t_w=2000\ s$ for the glass.  The
supercooled liquid curves have been averaged over the waiting-time
taking advantage of stationarity. The glassy curve has been smoothed
by averaging over two time-windows of length $\tau=200\ s$ 
centered at $t_w$ and $t$.

\begin{figure}
\begin{center}
\includegraphics[width=8cm]{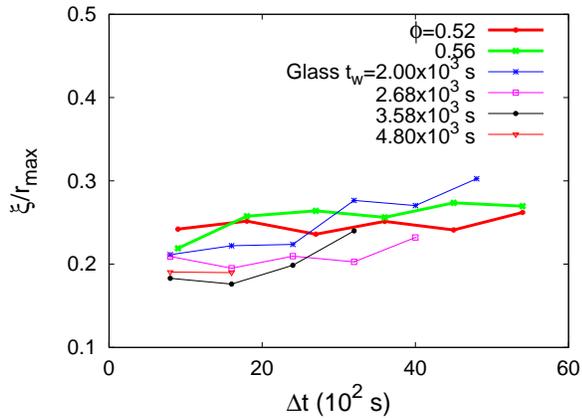}
\caption{\label{fig:xi} (Colour online) The time dependence of the
correlation length normalized by the cut-off distance 
$r_{max}$ ($=30$ lattice units) in the supercooled liquid and glass, computed
using eq.~(\ref{eq:xi}). 
}
\end{center}
\end{figure}

The correlation length can be evaluated using
\begin{equation}
\xi^2 \equiv \frac{\int_0^{r_{max}} dr \; r^2 \, C_4(r;,t,t_w)}
           {\int_0^{r_{max}}dr \; C_4(r;t,t_w)}
\; .
\label{eq:xi}
\end{equation}
in the limit $r_{max}\to \infty$. This analysis, applied to the data
in Fig.~\ref{fig:C4}, yields a correlation length of the order of $\xi
\sim 4-6R$. This can be confirmed by simple visual inspection since all
curves decay close to zero at distances $r\sim 8-10 \ \mu$m$\sim
4-6R$.  These values are of the same order as the ones found in
previous studies~\cite{Wecrwe}.  

In order to capture the
time-dependence of $\xi$ we use, instead, a finite value of $r_{max}$,
$r_{max}=30a \approx 7.1\ \mu$m. The reason is that 
we do not have enough precision at $r>r_{max}$ to disentangle the 
curves measured at different times. We thus obtain slightly shorter
values of $\xi$ that have, though, a systematic temporal trend.
Figure~\ref{fig:xi} shows
these results. The supercooled liquid curves follow the expected
trend: the sample with a higher packing fraction ($\phi\simeq 0.56$
with thick green line) has a longer correlation length than the one
with the lower packing fraction ($\phi\simeq 0.52$ with thick red
line). In both cases the length smoothly increases in time. We then
compare these results to the measurements in the glass at different
waiting-times (thin lines in the same figure).  All curves grow as a
function of time-difference. At short time-differences the curves with
shorter waiting-time have a longer correlation length while the trend
reverses at longer time-differences. In the glass the growth with
time-difference is faster than in the supercooled liquid and one
expects the longer waiting-time curves to go beyond the supercooled
liquid ones at longer time-differences (not reached in the
experiment). The reason why the correlation lengths in the glass are
shorter than the ones in the supercooled liquid at the available times
is that the glass is still far out of equilibrium and correlations
have not propagated far in the sample yet.

The two-time dependence in the glassy regime is similar to the one
found in the $3d$ EA spin-glass~\cite{3dEA}, the Lennard-Jones
mixture~\cite{Castillo}, and the $3d$ random field Ising coarsening
system~\cite{3dRFIM}. In the latter case the origin of the two-time
dependence of the growing length $\xi$ can be traced back to the
one-time dependence of the averaged radius of the growing domains of
two competing equilibrium states. In the structural and spin glass
cases, the two-time dependence of $\xi$ does not have such a clear
simple origin, and it is less well understood.

We now turn to the study of local correlations, which are probes of
local heterogeneities in the dynamics. The particles in the colloidal
system do not displace all at the same rate: some regions can
reconfigure much faster than others, for the same elapsed time between
frames. A broad distribution characterising these heterogeneities
can be captured, in the mapped spin system, by using a local two-time
spin-spin correlation averaged over a cell of size $V_r=\ell^3$
centered at site $r$:
\begin{equation}
C_\ell(r;t,t_w)=\frac{1}{V_r} \sum_{i=1 \in V_r}
s_i(t) s_i(t_w)
\;.
\end{equation} 
Whenever the
cell size $\ell$ is much larger than the dynamical correlation length
$\xi$, the local correlations just reflect the global value
$C_2(r=0;t,t_w)$. Instead, whenever the coarse graining box $\ell$ is
smaller than $\xi$, the local values are non-uniform. This fact is
captured by a broad probability distribution function (PDF) of the local
correlations $P(C_\ell)$ at fixed times $t$ and $t_w$. 

\begin{figure}
\begin{center}
\includegraphics[width=8cm]{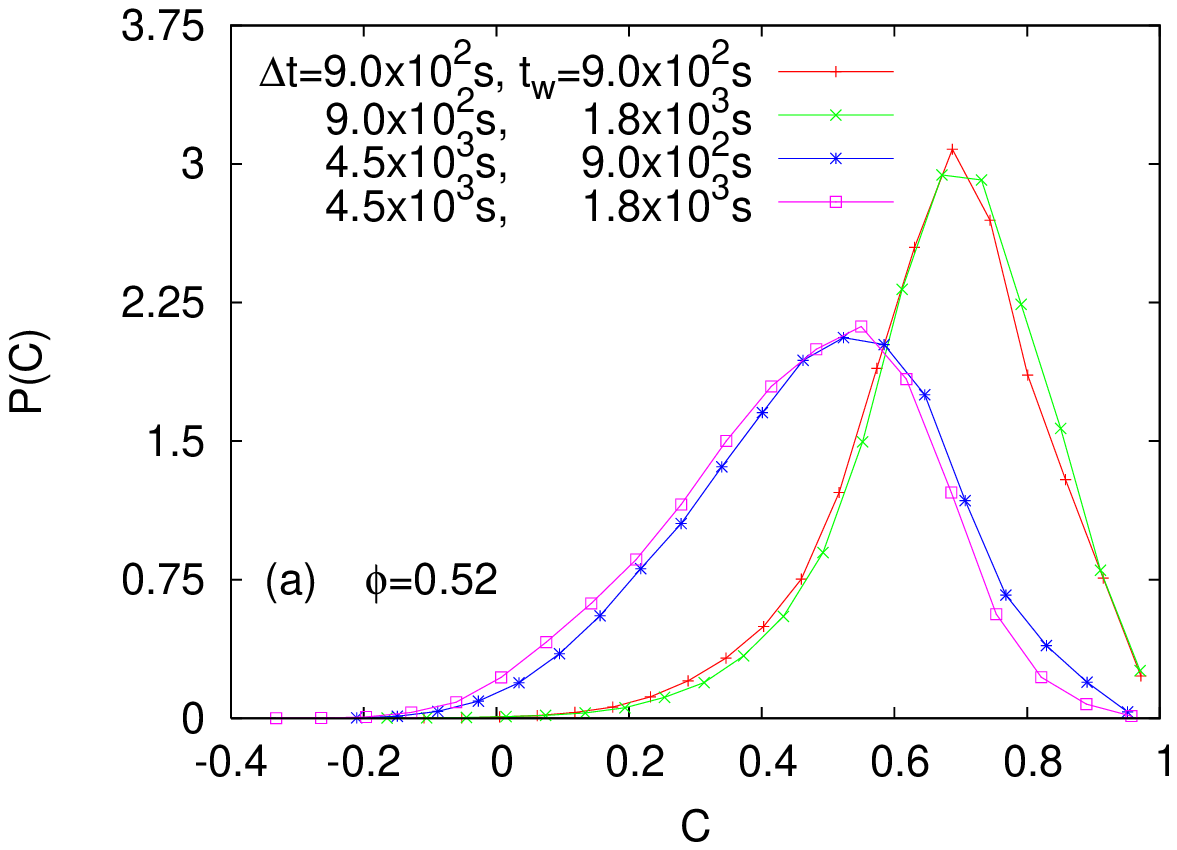}
\end{center}
\begin{center}
\includegraphics[width=8cm]{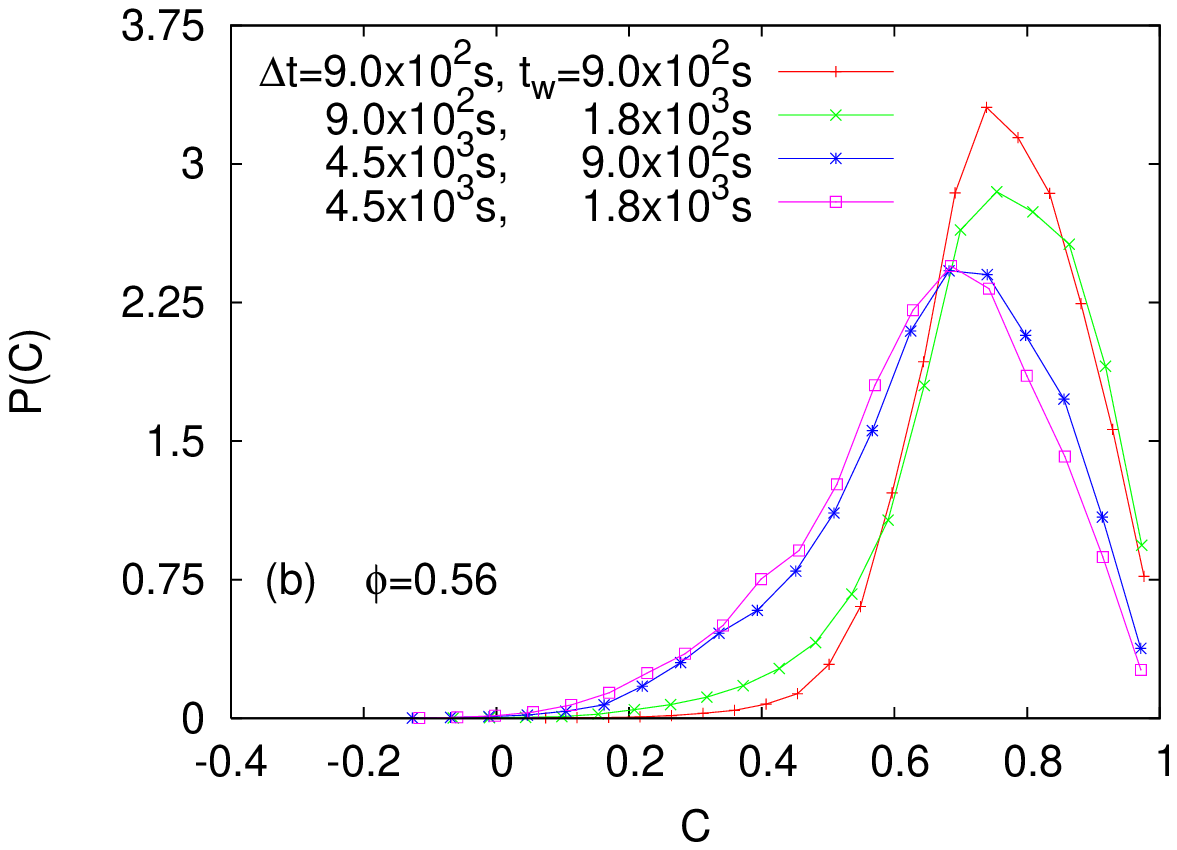}
\end{center}
\begin{center}
\includegraphics[width=8cm]{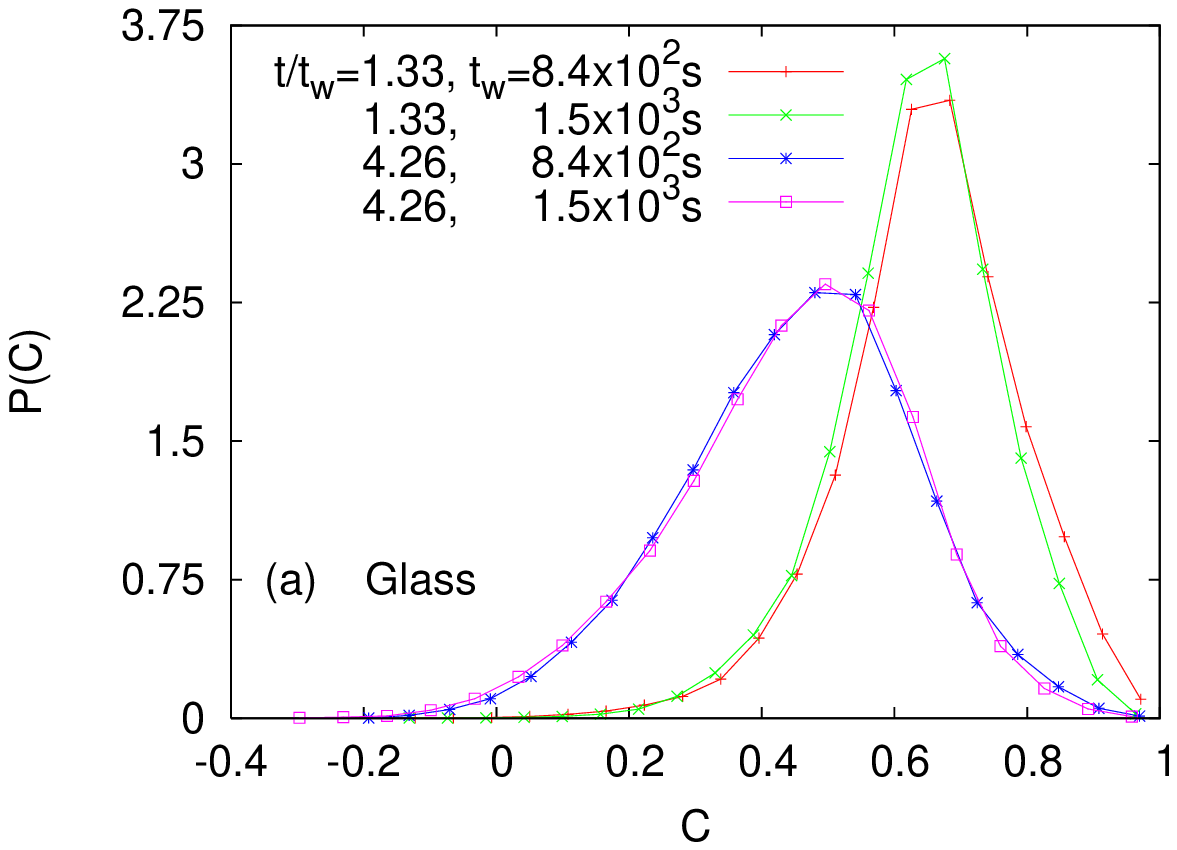}
\caption{\label{fig-local-pdfs} (Colour online) The PDFs for local
two-time spin-spin correlations for (a) supercooled liquid at
$\phi\simeq 0.52$, (b) supercooled liquid at $\phi\simeq 0.56$, and (c) the glass
at $\phi\simeq 0.62$. The distributions signal heterogeneous dynamics within
regions of linear size $\ell=2.2 R\simeq 2.60\ \mu m$. Supercooled
systems show data collapse at fixed time differences $t-t_w$,
reflecting time-translation invariance, while the glassy sample shows
data collapse at fixed $t/t_w$, reflecting approximate simple aging
behavior.}
\end{center}
\end{figure}

A simple scaling hypothesis discussed in \cite{3dEA,Chcu} implies
\begin{equation}
P(C_\ell; t,t_w,\ell,\phi) = P(C_\ell;C,\ell/\xi,\ell/L\to 0,\phi)
\end{equation}
with $C$ and $\xi$ the values of the global correlation and
correlation length at the measuring times $t$ and $t_w$, and $L$ the
size the of the sample that is much longer than the coarse graining
length. We kept an explicit dependence on the control parameter in the
system, that is to say, the packing fraction $\phi$.  This form is
obtained by exchanging the dependence on times $t$ and $t_w$ by a
dependence on the two-time dependence quantities $C$ and $\xi$
(exploiting their monotonicity properties) and then assuming that the
three lengths $L$, $\xi$ and $\ell$ can only appear through the ratios
$\ell/\xi$ and $\ell/L$.  Since the time-variation of the correlation
length is very slow, as a first approximation one can neglect the
scaling variable $\ell/\xi$ and simply check the scaling form
$P(C_\ell;C,\phi)$~\cite{Cachcuigke}.  In Fig.~\ref{fig-local-pdfs} we
test this scaling form using a coarse graining box $\ell=2.2 R\simeq
2.60\ \mu m$.  The PDFs are shown for different values of the waiting
time for the two supercooled systems at $\phi\simeq 0.52$ (a) and
$\phi\simeq 0.56$ (b), and for the glass with $\phi\simeq 0.62$
(c). The PDFs collapse for fixed time difference $\Delta t=t-t_w$ in
the case of the supercooled samples, and for fixed ratio $t/t_w$ in
the case of the glass, reflecting that time-translation invariance is
manifest in the supercooled liquid regime, but broken in the glass,
which ages.  (In the case of Fig.~\ref{fig-local-pdfs}(b) the bad
collapse for $t_w=900\ s$ may be attributed to lack of equilibration
at this high packing fraction: fluctuations may be more sensible than
average values in detecting a remanent time-variation. Another
explanation would be that the sample we are using is too much
heterogeneous and not fully representative of equilibration at this
packing fraction.)  As expected the PDFs get wider for longer
time-differences or larger value of $t/t_w$.

Remarkably, once the PDFs have been scaled, the scaling function of the
sample with low packing fraction is very similar to the one of the
glass. It is worth noting here that the average correlations in the
loose supercooled liquid and the glass are very similar: $\langle
C\rangle \sim 0.5$.  The averaged two-time correlation in the dense
supercooled liquid ($\phi\simeq 0.56$) remains, during the available
time-window, too high to be compared with the other two cases. The
similarity between the PDFs for $\phi\simeq 0.52$ and $0.62$ suggests that a
`universal' PDF connecting the fluctuations for different packing
fractions through a proper rescaling of times might exist. We plan to
explore this hypothesis using molecular dynamics of Lennard-Jones
mixtures.

\section{Conclusions}

In short, we introduced a simple method to translate particle data 
position into fixed frame spin variables. We computed correlation 
functions and extracted a correlation length from confocal microscopy
data of supercooled and glassy samples and we found remarkably similar
results to the ones obtained with numerical simulations of spin
models. 

One advantage
of our analysis method is that it does not rely on tracking
individual particles over time.  Instead, it acts on the particle
locations at each time, without regard to their identity.  Thus,
this method will be useful in situations where particles move too
rapidly to be tracked, but which are still able to be visualized by
microscopy in quickly obtained images.

\begin{acknowledgments}
LFC thanks the Universidad Nacional de Mar del Plata, Argentina, and
Boston University, and CC, GF and JLI thank the Laboratoire de
Physique Th\'eorique et Hautes Energies, Jussieu, France, for
hospitality during the preparation of this work.  The work of CC was
supported by the National Science Foundation under Grant
No. DMR-0403997.  The work of ERW was supported by the National
Science Foundation under Grant No. DMR-0239109.  LFC is a member of
Institut Universitaire de France. 
LFC, GF, and JLI acknowledge financial support from PICS 3172, 
PIP 5648 and PICT 20075.
\end{acknowledgments}



\begin{thebibliography}{20}
\bibitem{foot} Alternatively, one may work with fixed magnetization
and subtract this constant level from the spin variables. We choose to
work with the symmetric representation via the effective radius
$R_{\rm eff}$ so as to remain as close as possible to a spin glass
problem with zero magnetization, and thus make the analogies and
comparisons between the particle and spin systems easier and clearer.
\end{thebibliography}
\end{document}